\documentclass[11pt,letterpaper]{article}

\usepackage[left=1.12in,right=0.92in,top=0.86in,bottom=0.95in]{geometry}
\usepackage[T1]{fontenc}
\usepackage[utf8]{inputenc}
\usepackage{lmodern}
\usepackage{microtype}
\usepackage{graphicx}
\usepackage{xcolor}
\usepackage{booktabs}
\usepackage{tabularx}
\usepackage{array}
\usepackage{enumitem}
\usepackage{fancyhdr}
\usepackage{titlesec}
\usepackage{caption}
\usepackage{float}
\usepackage{amsmath}
\usepackage{tikz}
\usepackage[most]{tcolorbox}
\usepackage{listings}
\usepackage[numbers,sort&compress]{natbib}
\usepackage{hyperref}

\definecolor{RathBlue}{HTML}{2F80ED}
\definecolor{RathPink}{HTML}{EE5B95}
\definecolor{RathInk}{HTML}{141821}
\definecolor{RathMuted}{HTML}{667085}
\definecolor{RathLine}{HTML}{D9E2F0}
\definecolor{RathSoft}{HTML}{FBFCFE}
\definecolor{RathPinkSoft}{HTML}{FFF3F8}
\definecolor{RathCode}{HTML}{F8FAFC}
\definecolor{RathBlueMorandi}{HTML}{8FA6C9}
\definecolor{RathBlueMorandiSoft}{HTML}{F5F8FC}
\definecolor{RathPinkMorandi}{HTML}{C97A9E}
\definecolor{RathPinkMorandiSoft}{HTML}{FFF4F8}
\definecolor{RathMauveMorandi}{HTML}{A391B4}

\newcommand{\reportlogo}{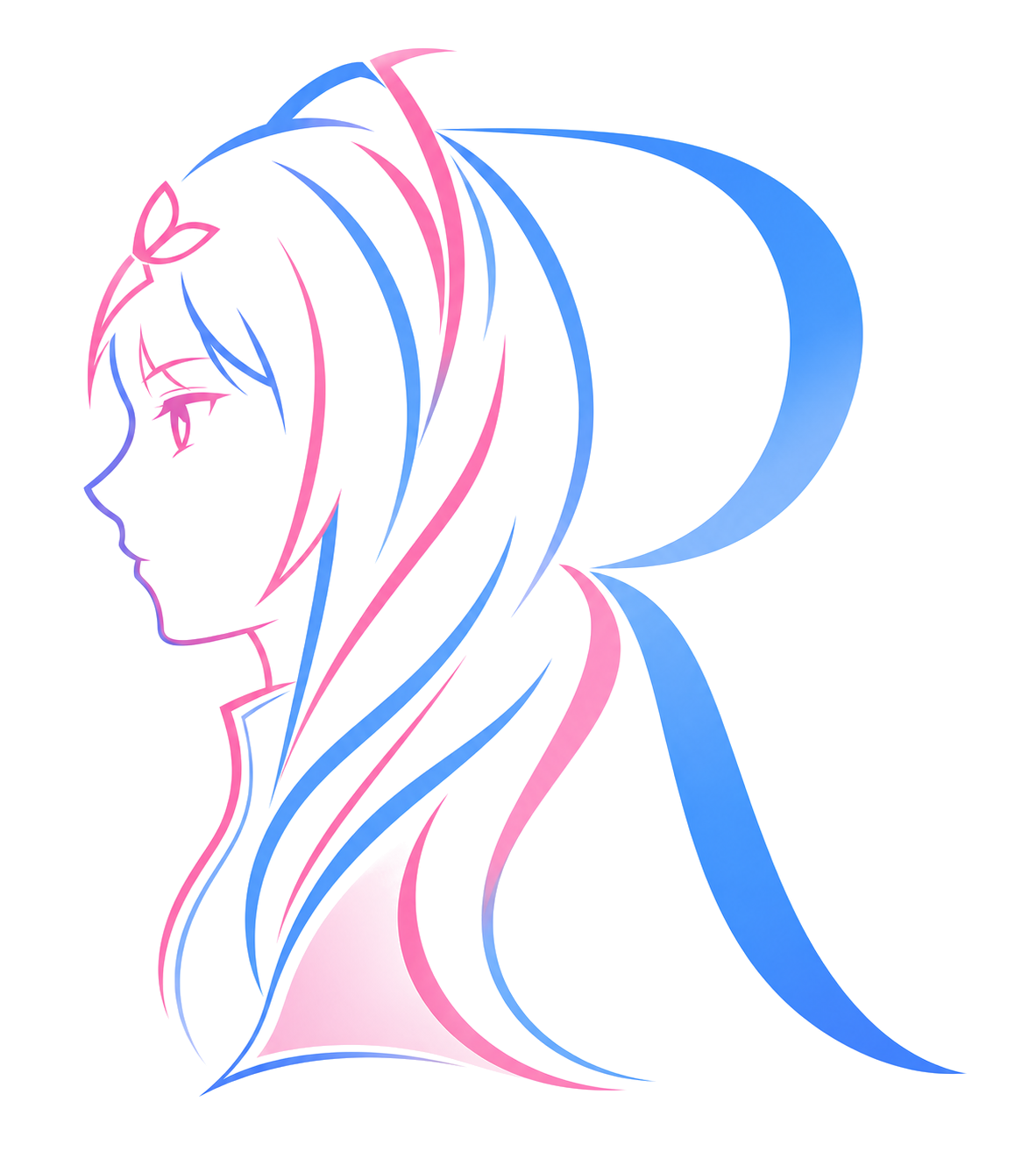}

\newcommand{\reportdisplaytitle}{OpenRath: Session-Centered Runtime State\\[-0.08em]for Agent Systems}
\newcommand{\reportheadtitle}{OpenRath: Session-Centered Runtime State}
\newcommand{\reportauthors}{Fukang Wen\textsuperscript{*,\textdagger}, Zhijie Wang\textsuperscript{*}, Ruilin Xu\textsuperscript{\textdagger}}

\newcommand{\reportcontact}{wfk25@mails.tsinghua.edu.cn}

\hypersetup{
  colorlinks=true,
  linkcolor=RathBlue,
  citecolor=RathBlue,
  urlcolor=RathBlue,
  pdftitle={OpenRath: Session-Centered Runtime State for Agent Systems},
  pdfauthor={Fukang Wen, Zhijie Wang, Ruilin Xu}
}

\setlist[itemize]{leftmargin=1.35em, itemsep=0.18em, topsep=0.2em}
\setlist[enumerate]{leftmargin=1.5em, itemsep=0.18em, topsep=0.2em}
\titleformat{\section}
  {\Large\sffamily\bfseries\color{RathInk}}
  {\textcolor{RathPink}{\thesection}}{0.65em}{}
\titleformat{\subsection}
  {\large\sffamily\bfseries\color{RathInk}}
  {\textcolor{RathMuted}{\thesubsection}}{0.65em}{}
\titlespacing*{\section}{0pt}{1.0em}{0.25em}
\titlespacing*{\subsection}{0pt}{0.72em}{0.18em}

\lstdefinestyle{rathpython}{
  language=Python,
  basicstyle=\ttfamily\footnotesize,
  keywordstyle=\color{RathBlue}\bfseries,
  commentstyle=\color{RathMuted},
  stringstyle=\color{RathPink},
  showstringspaces=false,
  breaklines=true,
  columns=fullflexible,
  frame=single,
  rulecolor=\color{RathLine},
  backgroundcolor=\color{RathCode},
  xleftmargin=0.5em,
  xrightmargin=0.5em,
  aboveskip=0.7em,
  belowskip=0.7em
}

\begin{document}

  \thispagestyle{fancy}
  \vspace*{-0.1em}
  \begin{center}
    {\LARGE\sffamily\bfseries\color{RathInk}\reportdisplaytitle\par}
    \vspace{0.42em}
    {\bfseries\reportauthors\par}
  \end{center}

  \vspace{0.28em}
  \begin{tcolorbox}[
    enhanced,
    width=\linewidth,
    colback=RathPinkSoft,
    colframe=RathPink,
    boxrule=0.36pt,
    arc=7pt,
    outer arc=7pt,
    left=18pt,
    right=18pt,
    top=11pt,
    bottom=10pt
  ]
    \begin{center}
      {\sffamily\bfseries\normalsize\color{RathPink}Abstract}
    \end{center}

Modern agent systems often suffer from fragmented runtime state: transcripts, tool effects, memory events, workspace placement, branch provenance, and replay evidence are recorded separately and become difficult to inspect or reproduce. OpenRath addresses this issue with a PyTorch-like programming model for multi-agent, multi-session systems. The analogy concerns the role of a central first-class runtime abstraction, not tensor computation. Its core abstraction is \texttt{Session}, the runtime value passed between agents and workflows.

A \texttt{Session} is branchable, inspectable, replayable, backend-aware, and composable. It records conversation chunks, sandbox placement, lineage metadata, token usage, pending work, and tool evidence, while defining where memory interactions enter the runtime record. Since this state is carried by the same value used in program execution, fork, merge, and replay become explicit runtime operations rather than states reconstructed from external traces.

OpenRath further defines \texttt{Sandbox}, \texttt{Tool}, \texttt{Agent}, \texttt{Memory}, \texttt{Workflow}, and \texttt{Selector}, with \texttt{Selector} turning control flow into runtime-routed decisions. This report presents the programming model, architecture, audited milestones, and evidence protocol. Its claims are limited to controlled runtime properties, while broad quantitative comparisons, live-provider quality, optional-backend availability, and memory quality are left for follow-on evaluation. The central thesis is that \texttt{Session} provides agent systems with a first-class runtime value for auditable composition.

  \end{tcolorbox}
  \begingroup
  \renewcommand{\thefootnote}{\fnsymbol{footnote}}
  \footnotetext[1]{Equal contribution: Fukang Wen and Zhijie Wang.}
  \footnotetext[2]{Corresponding authors: Fukang Wen and Ruilin Xu. Contact:
  \href{mailto:\reportcontact}{\reportcontact}.}
  \endgroup
  \vspace{0.72em}

\section{Introduction}

\subsection{Problem Framing}

Consider a long agent run: it plans, forks a branch to test an approach, calls tools, edits files inside a sandbox, recalls memory, compresses context, and eventually returns a correct answer. The natural audit questions are straightforward: which branch produced the final result, which tool modified which file, which memory item was recalled or committed, and which evidence was removed during compression? In many systems, the run cannot answer these questions. The final output may be correct, but the runtime state that produced it has been fragmented across side channels.

Modern agent applications increasingly resemble runtime systems rather than isolated conversations. A simple loop that interleaves reasoning and acting~\cite{react}---appending messages, calling a model, executing tools, and appending observations---remains a useful pattern for a single assistant. Yet this loop becomes a weak state boundary once work is distributed across roles, tools, memory stores, sandboxes, branches, and resumed executions.

OpenRath identifies this fragmentation as a hidden-runtime-state problem. A message list preserves the conversational surface, but it typically does not expose role provenance, abandoned branches, tool placement, workspace effects, memory recall or commit events, or evidence discarded by compression. Without lineage, tool evidence, sandbox metadata, and usage records, a final answer is difficult to audit once the model, provider, workspace, or prompt changes.

OpenRath therefore starts from the runtime-state boundary rather than from the number of agents in a loop. Its goal is to make the state passed between agents explicit enough to support composition, inspection, branching, merging, persistence, and evaluation. This is why the system is organized around \texttt{Session}. A \texttt{Session} is not merely chat history; it is the first-class runtime value that carries the evidence required to continue, review, and explain agent work.

\subsection{Central Claim}

The central claim is that agent systems benefit from a first-class runtime state, and OpenRath proposes \texttt{Session} as that state. To clarify the role of this object, OpenRath adopts a PyTorch-inspired programming model~\cite{pytorch}: not PyTorch's tensor mathematics, but its architectural interface for composable computation. In that interface, a central value flows through reusable modules, modules expose a uniform \texttt{forward} mapping, placement is made explicit through operations such as \texttt{tensor.to(device)}, and persistent module state is represented by parameters. OpenRath adapts this pattern to agent runtimes. \texttt{Session} plays the role of the flowing value; \texttt{Agent} is a reusable transformation similar in role to a layer; \texttt{Workflow} is a compositional container; both follow a \texttt{forward(session) -> session} contract; placement is expressed as \texttt{session.to(backend)}; and \texttt{Memory} is treated as an agent-bound persistent state plane rather than hidden prompt text.

Because each transformation preserves the \texttt{Session -> Session} shape, agents can be nested into workflows without introducing a separate runtime state format. Composition, branching, merging, handoff, and replay operate on ordinary program values rather than on state reconstructed from controller logs. The analogy is architectural rather than literal: the claim is not that agent systems are neural networks, but that agent runtimes need a stable flowing value, reusable transformations behind a uniform interface, explicit placement, persistent state, and inspectable evidence. The compact vocabulary that realizes this design---\texttt{Agent}, \texttt{Workflow}, \texttt{Tool}, \texttt{Memory}, and \texttt{Sandbox}---is developed in the programming model that follows.

Why make \texttt{Session} the runtime boundary, rather than placing this state inside a graph runtime's node state or a tracing system's spans? These layers serve different primary readers. Graph state records where execution is in a control flow, so that a run can resume, replay, or fork from checkpoints. Trace spans record what was observed during execution, such as model calls, tool calls, handoffs, guardrails, and other monitored events. Neither representation is designed to be the ordinary program value that agents themselves fork, merge, hand off, and replay. A trace is written for observers; a graph checkpoint is written for schedulers. A \texttt{Session} is written for the agent program: it is the live value passed through the program, and evidence is attached to that value rather than reconstructed from a side channel. OpenRath's design hypothesis is that multi-agent systems remain more inspectable as they scale when runtime state is placed where the program already flows, rather than beside it.

\begin{table}[!htbp] \vspace{0.1em} \centering \small \caption{Three runtime records, three readers. OpenRath's \texttt{Session} is the value written for the agent program itself, which is why fork, merge, and replay are first-class rather than reconstructed.} \label{tab:session-vs-others} \begin{tabularx}{\linewidth}{@{}>{\raggedright\arraybackslash}p{0.22\linewidth}>{\raggedright\arraybackslash}X>{\raggedright\arraybackslash}X@{}} \toprule Record & Written for & What it primarily holds \\ \midrule Graph checkpoint & The scheduler & Where execution is in the control flow, so a run can resume or time-travel. \\ Trace span & The observer & What was observed during a run, for monitoring and debugging after the fact. \\ \texttt{Session} (OpenRath) & The agent program & The live value agents fork, merge, hand off, and replay; lineage, tools, placement, memory, and usage travel with it. \\ \bottomrule \end{tabularx} \vspace{0.1em} \end{table}

\subsection{Ecosystem Positioning}

The ecosystem positioning is intentionally narrow. Agent infrastructure is
moving toward durable execution, richer tracing, standardized tool/data
protocols, and real-environment evaluation; representative systems include
AutoGen~\cite{autogen}, LangGraph~\cite{langgraph-persistence}, the OpenAI
Agents SDK~\cite{openai-agents-tracing}, and MCP~\cite{mcp-docs,anthropic-mcp}.
OpenRath complements those layers by working at a different boundary: the
runtime value that carries their effects. A graph runtime can schedule work, a
tracing system can observe spans, MCP can expose tools, and a sandbox can run
commands; OpenRath asks how those effects become one branchable, inspectable,
replayable state object that agent programs can pass between agents and
workflows. Related Work develops this layer by layer; here it is enough to fix
the boundary.

OpenRath is neither a universal substitute for graph runtimes, tracing systems, MCP servers, sandbox providers, or benchmark harnesses, nor a thin wrapper around any one of them. The intended role is connective: a \texttt{Session} is the object that can
be scheduled, traced, dispatched, persisted, forked, merged, compressed, and
reviewed without forcing each layer to invent its own incompatible
representation of agent state. The central claim is smaller and more defensible:
multi-agent systems need a first-class runtime state, and \texttt{Session} is
OpenRath's candidate for that state.

\subsection{Contributions}

This report makes four technical contributions to the runtime-state boundary
for agent systems.

\begin{enumerate}[itemsep=0.1em, topsep=0.2em]
  \item \textbf{A session-centered runtime dataflow.}
  OpenRath treats \texttt{Session} as the value that moves through the agent
  runtime, so conversation chunks, placement, lineage, usage, pending work,
  tool evidence, and memory-boundary records are represented as one inspectable
  flow rather than separate controller bookkeeping.

  \item \textbf{A PyTorch-like object vocabulary for agent programs.}
  The framework organizes agent programs around
  \texttt{Session}, \texttt{Sandbox}, \texttt{Tool}, \texttt{Agent},
  \texttt{Memory}, \texttt{Workflow}, and \texttt{Selector}. Each object has a
  narrow runtime boundary while preserving the same \texttt{Session -> Session}
  shape, including runtime-routed control flow.

  \item \textbf{Backend-aware boundaries for tools and memory.}
  OpenRath separates runtime state from the execution backend that runs tools
  and the memory backend that persists recallable state. This lets local
  execution, optional OpenSandbox placement, MCP-style tools, and memory
  services participate in one session-centered model as their evidence packets
  are verified.

  \item \textbf{An audit-first release protocol.}
  The report maps claims to packets: lineage export, local sandbox execution,
  workflow transcript, focused tests, visual QA, claim ledger, and a
  memory source audit. Broad benchmark superiority, human preference
  results, and cross-system leaderboard claims are reserved for follow-on
  quantitative evaluation.
\end{enumerate}

\subsection{Runtime State at a Glance}
The core visual distinction is simple. A loop-centered agent treats messages,
tool logs, memory updates, usage, workspace effects, and branch provenance as
side channels around the loop. OpenRath moves those effects into one typed
runtime value. The same \texttt{Session} can be passed to agents, forked for
independent work, merged after review, persisted as evidence, and replayed with
explicit backend boundaries. OpenRath does not replace tool protocols, sandbox
providers, memory stores, tracing systems, or graph schedulers. It records
their effects in a session object that can move through the program as
branchable, inspectable, and replayable runtime state.

\begin{figure}[!htbp]
\centering
\includegraphics[width=0.96\linewidth]{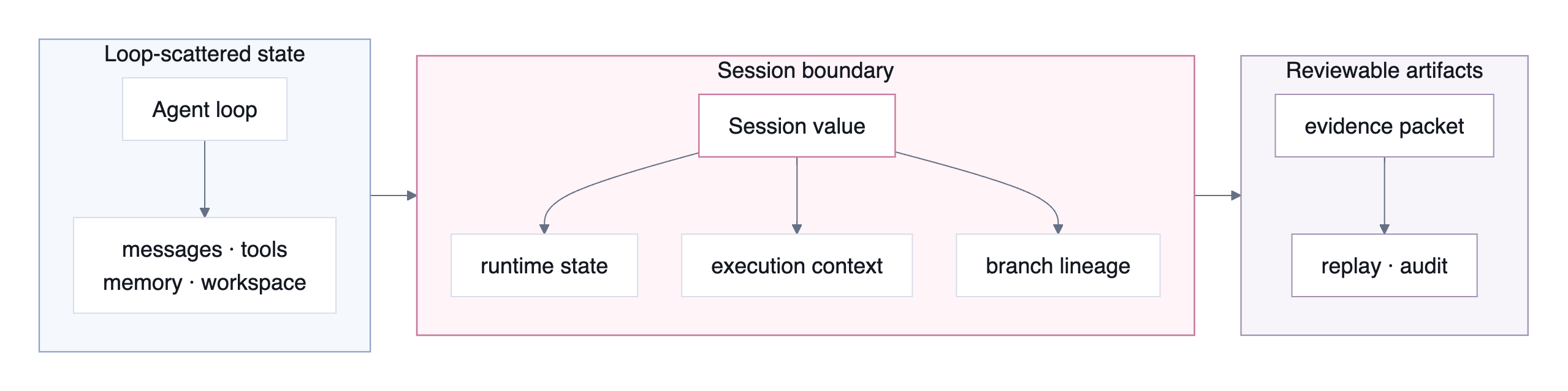}
\caption{OpenRath's core boundary: side-channel state around an agent loop is
promoted into a branchable \texttt{Session} value that can produce release
evidence artifacts.}
\label{fig:session-runtime-boundary}
\end{figure}

\section{Related Work}

The agent ecosystem is converging on a specialized runtime stack: reasoning-and-acting
methods, multi-agent frameworks, durable graph runtimes, tracing SDKs, tool/data
protocols, real-environment benchmarks, and provenance standards each own one
layer. The open design question is the \emph{crossing object}---what state can
move through these layers while keeping conversation, lineage, placement, tool
effects, memory, and artifacts together. We survey these areas by that question
and, for each, mark the distinction from OpenRath's \texttt{Session}.

\subsection{Tool-Using and Acting Agents}

Chain-of-thought prompting and self-consistency elicit and stabilize
intermediate reasoning at inference time~\cite{cot,selfconsistency}. ReAct
interleaves reasoning with environment-directed actions~\cite{react}, and MRKL
combines a model with external knowledge and discrete reasoning
modules~\cite{mrkl}. Tool use itself is taught or routed by
Toolformer~\cite{toolformer}, HuggingGPT's controller over expert
models~\cite{hugginggpt}, and Gorilla's retrieval-grounded API
calls~\cite{gorilla}, and is studied at scale by ToolLLM, API-Bank, and
ToolAlpaca~\cite{toolllm,apibank,toolalpaca}; Tree of Thoughts adds deliberate
search with backtracking~\cite{tot}, an inference-time search rather than a
persistent, replayable branch. These works advance \emph{how} a model reasons
and acts; OpenRath is complementary, making the runtime state those actions
produce---lineage, tool evidence, placement---a first-class value.

\subsection{Multi-Agent Frameworks}

AutoGen frames applications as multi-agent conversations~\cite{autogen}, CAMEL
studies role-playing communicative agents~\cite{camel}, MetaGPT encodes
standardized operating procedures into a collaboration pipeline~\cite{metagpt},
ChatDev runs a virtual software company over a chat chain~\cite{chatdev}, and
AgentVerse studies dynamic group collaboration and emergent
behavior~\cite{agentverse}. These contribute orchestration patterns. The
distinction is one of object boundary: where MetaGPT's SOP governs \emph{which
role acts when}, OpenRath governs \emph{what value the roles pass}, so
multi-agent composition needs no second, framework-private state object.

\subsection{Runtime State, Protocols, and Observability}

LangGraph exposes checkpointed graph state with history and time travel to
replay or fork from a checkpoint~\cite{langgraph-persistence,langgraph-timetravel};
an OpenRath \texttt{Session}, by contrast, is the value the program itself passes
and forks, not a scheduler checkpoint. The OpenAI Agents SDK records traces and
spans over generations, tool calls, handoffs, and guardrails and composes agents
from those parts~\cite{openai-agents-tracing,openai-agents-docs}, and
OpenTelemetry treats spans as an observer-facing signal~\cite{opentelemetry};
traces describe \emph{what was observed} after the fact, whereas \texttt{Session}
is written for the program, so its evidence is the value itself. Connectivity is
standardized by the Model Context Protocol~\cite{mcp-docs,anthropic-mcp} and
interface descriptions such as OpenAPI~\cite{openapi}. The dataflow-runtime
analogy is instructive: TensorFlow represents computation and shared state as a
graph~\cite{tensorflow}, while OpenRath keeps the value imperative and lets
lineage, placement, and evidence travel with it.
Table~\ref{tab:related-runtime-landscape} summarizes the object-boundary
question each layer leaves open.

\begin{figure}[!htbp]
\centering
\includegraphics[width=0.96\linewidth]{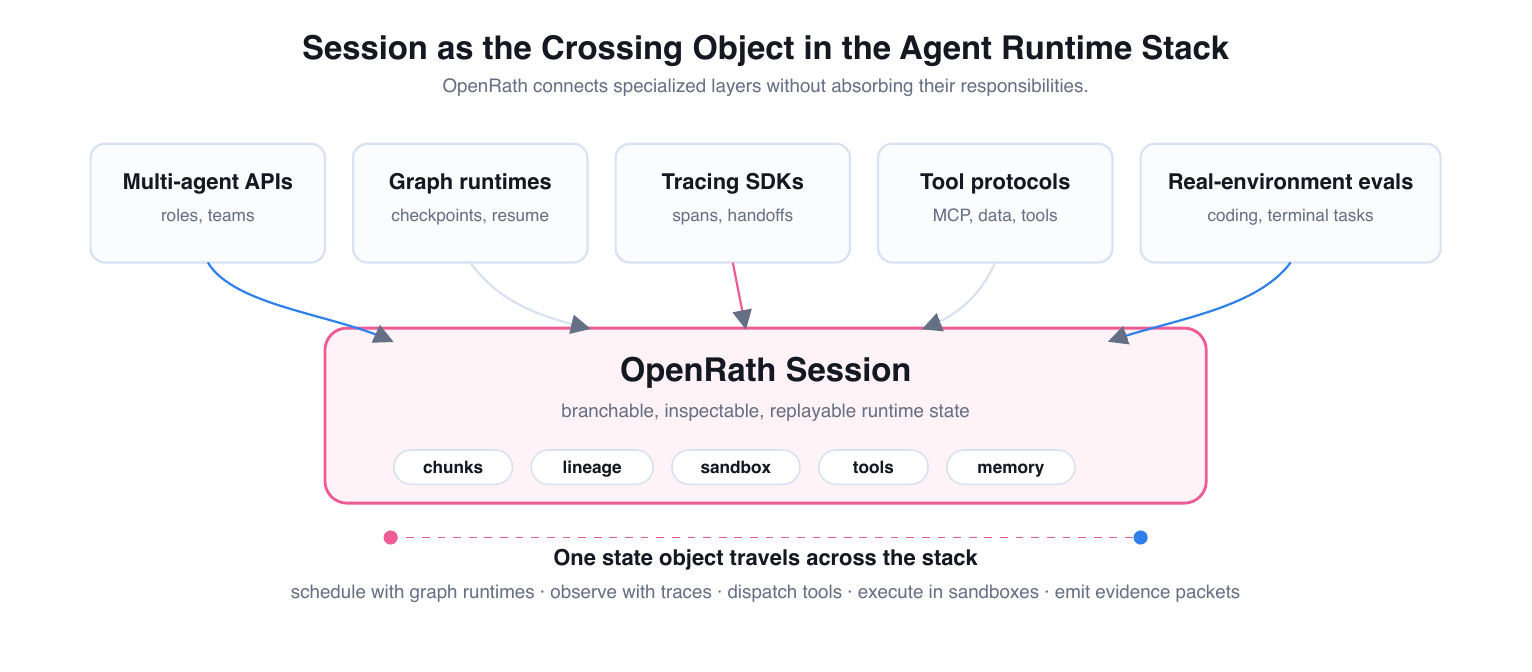}
\caption{OpenRath's ecosystem role is a crossing-object boundary. It can work
with specialized agent APIs, graph runtimes, tracing SDKs, tool protocols,
sandbox providers, and evaluation harnesses by making their effects visible in
one \texttt{Session}.}
\label{fig:runtime-stack-positioning}
\end{figure}

\begin{table}[H]
\vspace{0.2em}
\centering
\caption{Runtime-stack trends and OpenRath's intended boundary.}
\label{tab:related-runtime-landscape}
\begin{tabularx}{\linewidth}{@{}>{\raggedright\arraybackslash}p{0.23\linewidth}>{\raggedright\arraybackslash}X>{\raggedright\arraybackslash}X@{}}
\toprule
Trend & What became first-class & Remaining object-boundary question \\
\midrule
Multi-agent APIs & Agent roles, teams, group chats, and workflow patterns. & What state moves between roles besides transcript text? \\
Durable graph runtimes & Checkpoints, thread state, state history, resume, and time travel. & What session-level evidence should a graph node carry? \\
Tracing SDKs & Model spans, tool calls, handoffs, guardrails, and custom events. & What runtime value should traces attach to and replay from? \\
Tool/data protocols & Standardized access to external tools, data, and workflows. & How do external effects return as lineage, artifacts, and backend evidence? \\
Real-environment benchmarks & Repository tasks, terminal tasks, GUI/CLI environments, tests, and scored outcomes. & Can reviewers inspect how the outcome was produced? \\
\bottomrule
\end{tabularx}
\vspace{0.2em}
\end{table}

\subsection{Memory and Retrieval}

A large body of work studies what an agent should remember and how it should
retrieve it. At the agent level, Reflexion converts feedback from failed
attempts into natural-language reflections held in an episodic buffer, so later
attempts improve~\cite{reflexion}; Generative Agents maintain a long-running
memory stream that is retrieved, reflected upon, and compiled into
plans~\cite{generative-agents}; and MemGPT adopts the operating-system idea of a
memory hierarchy, paging information between a bounded context window and
external storage to sustain long interactions~\cite{memgpt}. Voyager carries
this toward lifelong skill acquisition, growing a reusable library of verified
behaviors from environment feedback~\cite{voyager}. OpenRath does not propose a new memory model; it simply makes memory operations session-visible, so recall and commit are recorded as explicit runtime events on \texttt{Session} rather than hidden inside the prompt.

\subsection{Agent Benchmarks and Environments}

Interactive evaluation spans AgentBench~\cite{agentbench} and $\tau$-bench's
tool-agent-user setting with database-state checks~\cite{taubench}; software
engineering through SWE-bench~\cite{swebench}, SWE-agent's agent-computer
interface~\cite{swe-agent}, and the human-filtered SWE-bench Verified
subset~\cite{swebench-verified}; terminals through Terminal-Bench, TerminalWorld,
and task-alignment studies~\cite{terminal-bench,terminalworld,tab-benchmark}; and
web, desktop, and embodied settings including WebArena, VisualWebArena,
WorkArena, OSWorld, WebShop, Mind2Web, ALFWorld, ScienceWorld, GAIA, and
TheAgentCompany~\cite{webarena,visualwebarena,workarena,osworld,webshop,mind2web,alfworld,scienceworld,gaia,theagentcompany}.
These score outcomes inside realistic environments; OpenRath's complementary
question is whether the trajectory that produced an outcome is inspectable and
replayable---a precondition for trustworthy scoring.

\section{Background and Motivation}

Section~1 framed the gap with a single run, and the related work above situated
OpenRath among adjacent agent systems---tool-using and acting agents, multi-agent
frameworks, runtime/observability layers, memory and retrieval, evaluation
environments, and provenance standards. This section states the gap that runs
underneath all of them as a general property of multi-agent work. The loop
boundary that suffices for one assistant becomes a weak state boundary the moment
an application branches across roles, tools, memory, files, sandboxes, and
resumed runs. The transcript still shows a final answer, but the runtime path
that produced it is spread across controller code, tool logs, memory stores,
workspace state, and provider traces.

This pressure is becoming more visible as agent products move from demos to
longer-running workflows. Once an agent edits a repository, calls external
tools, resumes after interruption, or routes work through multiple roles, the
state boundary becomes an engineering contract rather than an implementation
detail. Users and reviewers need to ask ordinary runtime questions: what was
the input state, what changed, which backend performed the change, which memory
or artifact influenced the decision, and how can the run be resumed or
replayed? A framework that cannot answer those questions may still produce a
plausible response, but it cannot easily support release review, debugging,
audit, or systematic evaluation.

This hidden state is the central motivation for OpenRath. Multi-agent work is
not merely one conversation with more roles. It naturally creates multiple
runtime paths: one branch gathers context, another edits or tests an artifact,
another validates evidence, and another compresses the result. If every
intermediate step is placed into one shared transcript, later agents inherit too
much noise. If intermediate work is hidden in controller state, reviewers cannot
reconstruct which branch produced a claim, what memory was recalled, which
sandbox touched a file, or what evidence was discarded during compression.

OpenRath treats branchability as a property of runtime state rather than as a controller-side convention. A branch should inherit the portion of parent context required for independent work, accumulate local evidence during execution, and merge useful results back without erasing provenance. The object being branched is therefore \texttt{Session}: the runtime value that flows through agents, tools, memory-boundary operations, sandbox placement, compressors, and workflows.

This boundary is the foundation for the remainder of the report. The next section makes it concrete through a compact object vocabulary centered on \texttt{Session}, the value that every other component reads, transforms, annotates, or passes forward.

\section{OpenRath Programming Model}

OpenRath keeps the programming model small on purpose. The core rule is that
runtime components transform or annotate \texttt{Session}; they should not each
invent a private transcript, placement record, tool log, memory format, or
workflow state. This rule is what makes the PyTorch analogy useful: the analogy
is not about tensor math, but about one value flowing through reusable
transformations with explicit placement and persistent state boundaries.

\begin{table}[H]
\vspace{0.15em}
\centering
\small
\caption{The compact OpenRath object vocabulary.}
\label{tab:programming-model-vocabulary}
\begin{tabularx}{\linewidth}{@{}lX@{}}
\toprule
Object & Runtime boundary \\
\midrule
\texttt{Session} & Flowing runtime value for chunks, placement, lineage, usage, pending work, tool evidence, and memory evidence when enabled. \\
\texttt{Agent} & Reusable \texttt{Session -> Session} transformation with local prompt, provider, tools, and memory policy. \\
\texttt{Tool} & Model-visible callable operation backed by schema validation, session context, sandbox dispatch, and returned evidence. \\
\texttt{Sandbox} & Placement boundary for file, command, code, and external tool execution. \\
\texttt{Memory} & Intended persistent-state plane for recall and commit across runs, kept separate from prompt text. \\
\texttt{Workflow} & Composition surface for agents, tools, branches, compression, memory, and child workflows. \\
\texttt{Selector} & Runtime router over self-describing workflows: it reads the current session and picks the next workflow, so dynamic control flow stays explicit instead of hard-coded. \\
\bottomrule
\end{tabularx}
\vspace{0.15em}
\end{table}

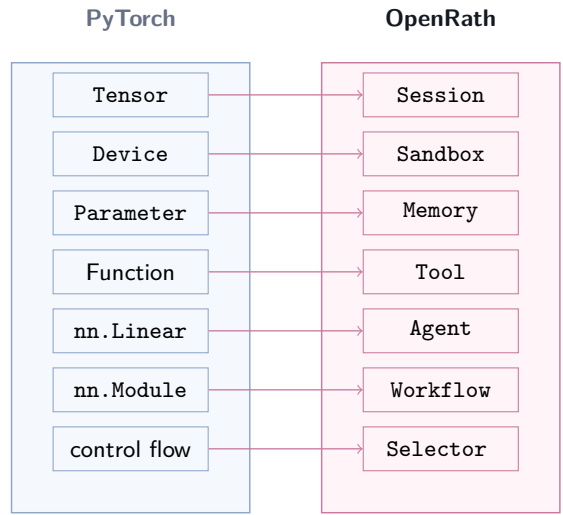
\begin{figure}[!htbp]
\centering
\begin{tikzpicture}[
  font=\footnotesize\sffamily,
  leftbox/.style={draw=RathBlueMorandi, fill=RathBlueMorandiSoft, minimum width=2.05cm, minimum height=0.58cm, align=center, inner sep=2pt},
  rightbox/.style={draw=RathPinkMorandi, fill=RathPinkMorandiSoft, minimum width=2.05cm, minimum height=0.58cm, align=center, inner sep=2pt},
  pt/.style={text=RathMuted},
  ra/.style={->, draw=RathPinkMorandi, line width=0.5pt}
]
\node[draw=RathBlueMorandi, fill=RathBlueMorandiSoft, line width=0.55pt, minimum width=3.15cm, minimum height=5.95cm] at (0, -2.55) {};
\node[draw=RathPinkMorandi, fill=RathPinkMorandiSoft, line width=0.55pt, minimum width=3.15cm, minimum height=5.95cm] at (4.1, -2.55) {};
\foreach \i/\ptlabel/\orlabel in {
  0/{\texttt{Tensor}}/{\texttt{Session}},
  1/{\texttt{Device}}/{\texttt{Sandbox}},
  2/{\texttt{Parameter}}/{\texttt{Memory}},
  3/{Function}/{\texttt{Tool}},
  4/{\texttt{nn.Linear}}/{\texttt{Agent}},
  5/{\texttt{nn.Module}}/{\texttt{Workflow}},
  6/{control flow}/{\texttt{Selector}}
}{
  \node[leftbox] (p\i) at (0, -\i*0.78) {\ptlabel};
  \node[rightbox] (o\i) at (4.1, -\i*0.78) {\orlabel};
  \draw[ra] (p\i.east) -- (o\i.west);
}
\node[pt, anchor=south, font=\footnotesize\sffamily\bfseries] at (0, 0.72) {PyTorch};
\node[anchor=south, text=RathInk, font=\footnotesize\sffamily\bfseries] at (4.1, 0.72) {OpenRath};
\end{tikzpicture}
\caption{The PyTorch lens. Each agent-runtime concern maps onto one OpenRath
object, with \texttt{Session} as the flowing value (the tensor of the runtime)
and \texttt{Selector} routing control flow at run time. The mapping is a
teaching device, not a claim that agent systems are neural networks.}
\label{fig:pytorch-lens}
\end{figure}

The most important design choice is what each object does not own. An
\texttt{Agent} does not own the entire conversation graph; lineage belongs to
\texttt{Session}. A \texttt{Tool} does not own placement; it executes through
the active sandbox. A \texttt{Workflow} does not create a separate orchestration
state; it composes transformations over sessions. \texttt{Memory} does not
become hidden prompt text; recall and commit should remain visible runtime
events. These separations keep the system inspectable when a run becomes
multi-agent, multi-branch, and multi-backend.

The tool boundary illustrates the pattern. A flow-level tool exposes a name,
description, and JSON schema to the model, while its Python call receives the
active \texttt{Session} and validated arguments. Built-in tools can then create
backend payloads for file, command, code, or MCP-like execution without changing
the model-visible contract. The same principle applies to workflows: a workflow
may fork a session, call an agent, validate in a sandbox, compress context, and
return a new session, but the evidence remains attached to the shared runtime
value.

Control flow follows the same discipline through \texttt{Selector}. Rather than
hard-coding which agent or workflow runs next, a \texttt{Selector} reads the
current \texttt{Session} and routes to one of several self-describing
workflows, returning an empty workflow when the task is done. This keeps
branching and looping over agents as ordinary, inspectable runtime decisions:
the routing choice becomes part of the session record instead of vanishing into
controller code. It is also where OpenRath departs from a static workflow
graph---the next step is decided at runtime from session state, yet every
decision still flows through one value.

Memory is described with a deliberately bounded claim. OpenRath provides local
memory with lexical recall, optional embeddings, and an optional external
backend, exposed through agent-level recall and commit operations so that
remembering and recalling stay visible runtime events rather than hidden prompt
text. What this report does not claim is retrieval \emph{quality}: how well a
given corpus, embedding choice, and commit policy serve a task is an empirical
question left to a follow-on evaluation. The programming model reserves the
correct boundary---memory as a session-visible persistent plane---without
asserting that every quality and backend trade-off has been measured.

\section{Runtime Architecture}

The runtime architecture answers one question: how does a \texttt{Session}
remain inspectable as it moves through agents, tools, sandboxes, branches, and
stored artifacts? OpenRath uses a small lifecycle rather than a separate runtime
object for every phase. A session is created from user or agent context, placed
on an execution backend when needed, transformed by agents or workflows,
branched for parallel work, merged after review, persisted for replay, and
released when sandbox resources are no longer owned.

\begin{figure}[!htbp]
\centering
\includegraphics[width=0.94\linewidth]{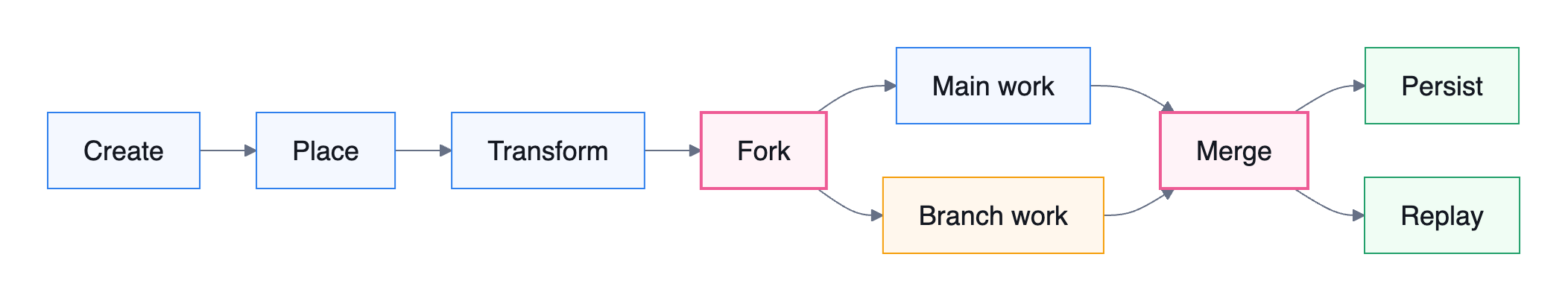}
\caption{Session lifecycle as a single runtime value: the same object is placed,
transformed, branched, merged, persisted, and replayed rather than replaced by a
separate orchestration state.}
\label{fig:session-lifecycle}
\end{figure}

\begin{table}[!htbp]
\vspace{0.15em}
\centering
\small
\caption{Runtime path of an OpenRath session.}
\label{tab:runtime-session-path}
\begin{tabularx}{\linewidth}{@{}lX@{}}
\toprule
Phase & What becomes auditable \\
\midrule
Create & Initial user text, agent prompts, role, and ordered chunk state. \\
Place & Backend intent from \texttt{Session.to(...)} and the workspace where tools can run. \\
Transform & Model calls, tool requests, results, errors, compressors, workflow steps, and usage. \\
Branch & Parent sessions, \texttt{fork}, \texttt{detach}, \texttt{merge}, branch provenance, and merge inputs. \\
Persist & Replayable chunks, lineage JSONL rows, usage, and source evidence. \\
Release & Sandbox handle ownership and backend lifetime. \\
\bottomrule
\end{tabularx}
\vspace{0.15em}
\end{table}

Branching is the point where a transcript becomes a graph. \texttt{fork}
duplicates state while preserving the parent relation; \texttt{detach} starts a
new lineage root from copied content; \texttt{merge} joins compatible sessions
and records both parents. In the current implementation, merge compatibility
includes sandbox compatibility: sessions must share a live sandbox handle or
target the same unbound backend. This makes placement part of the runtime graph
rather than an external execution detail.

Tool execution follows a layered path. The model sees FlowToolCall schemas. The session loop
combines built-in and user tools, sends schemas to the provider, resolves returned tool calls by
name, validates arguments, and invokes the selected tool with the active session. When a tool
needs side effects, it dispatches a backend payload through the session’s sandbox. Malformed
arguments, unknown tools, exceptions, and successful results all become tool-result chunks rather
than disappearing into controller flow.

\begin{figure}[!htbp]
\centering
\includegraphics[width=0.94\linewidth]{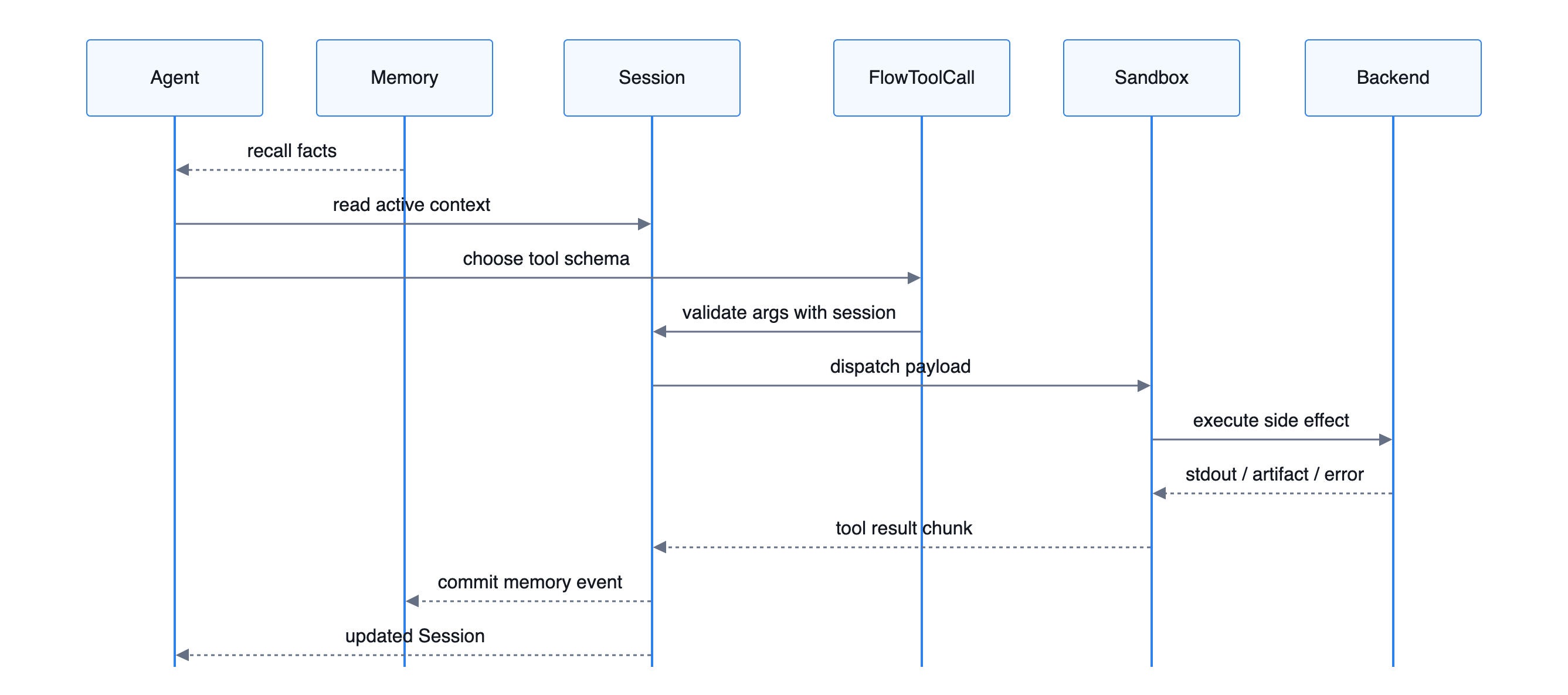}
\caption{Tool execution boundary: schemas are visible to the model, side effects
run through the session's sandbox and backend, and results return as session
evidence.}
\label{fig:tool-execution-sequence}
\end{figure}

\begin{table}[!htbp]
\vspace{0.15em}
\centering
\small
\caption{Backend boundary used by tool execution.}
\label{tab:runtime-backend-boundary}
\begin{tabularx}{\linewidth}{@{}>{\raggedright\arraybackslash}p{0.25\linewidth}>{\raggedright\arraybackslash}X@{}}
\toprule
Boundary & Owner and role \\
\midrule
Placement intent & \texttt{Session} stores the backend name and opening spec before execution. \\
Resource lifetime & A sandbox handle shares or releases live execution resources across branches. \\
Capability claim & The backend class advertises isolation level and supported tool payloads. \\
Concrete execution & The backend instance runs local, OpenSandbox, or future backend operations. \\
Evidence return & The session loop appends results or errors to the session stream. \\
\bottomrule
\end{tabularx}
\vspace{0.15em}
\end{table}

Persistence and replay close the loop. A running session can append rows to its
session JSONL store; lineage export can project sessions into plain JSONL rows
containing identifiers, parent identifiers, lineage operator, lineage kind,
chunk count, and cumulative usage. The format is intentionally boring: it can be
inspected with command-line tools, attached to release evidence, and converted
into diagrams later. This is the architectural through-line of the report:
OpenRath makes agent work easier to evaluate because conversation, tools,
placement, lineage, usage, and replay artifacts are carried by the same
session-centered runtime path.

\section{Multi-Agent Multi-Session Design}

Multi-agent design in OpenRath is intentionally small: an agent is a reusable layer, a workflow
is a reusable composition, and the moving runtime value is still Session. This avoids a common
failure mode in agent systems, where a single-agent API works cleanly but the multi-agent version
introduces a new shared mutable object, a hidden message bus, or a controller-only trace.

The engineering examples use this shape for lead-engineer, specialist, and QA roles; the research
examples use the same shape for literature, reproduction, compression, and output stages. The
domain-specific roles differ, but the runtime contract does not. This is the point of the design: a
workflow can grow from a script into a nested agent team without replacing the object that carries
evidence, placement, lineage, usage, and replay state.

This report therefore treats multi-agent capability as a runtime-state claim, not as a claim that
every workflow is already a measured benchmark result. Current evidence verifies deterministic
lineage export, local sandbox packets, workflow transcripts, focused tests, and layout review. Larger
claims about parallel branch scheduling, merge quality, memory quality, and task-level leaderboards remain scoped to follow-on evaluation.

\begin{figure}[!htbp]
\centering
\includegraphics[width=0.98\linewidth]{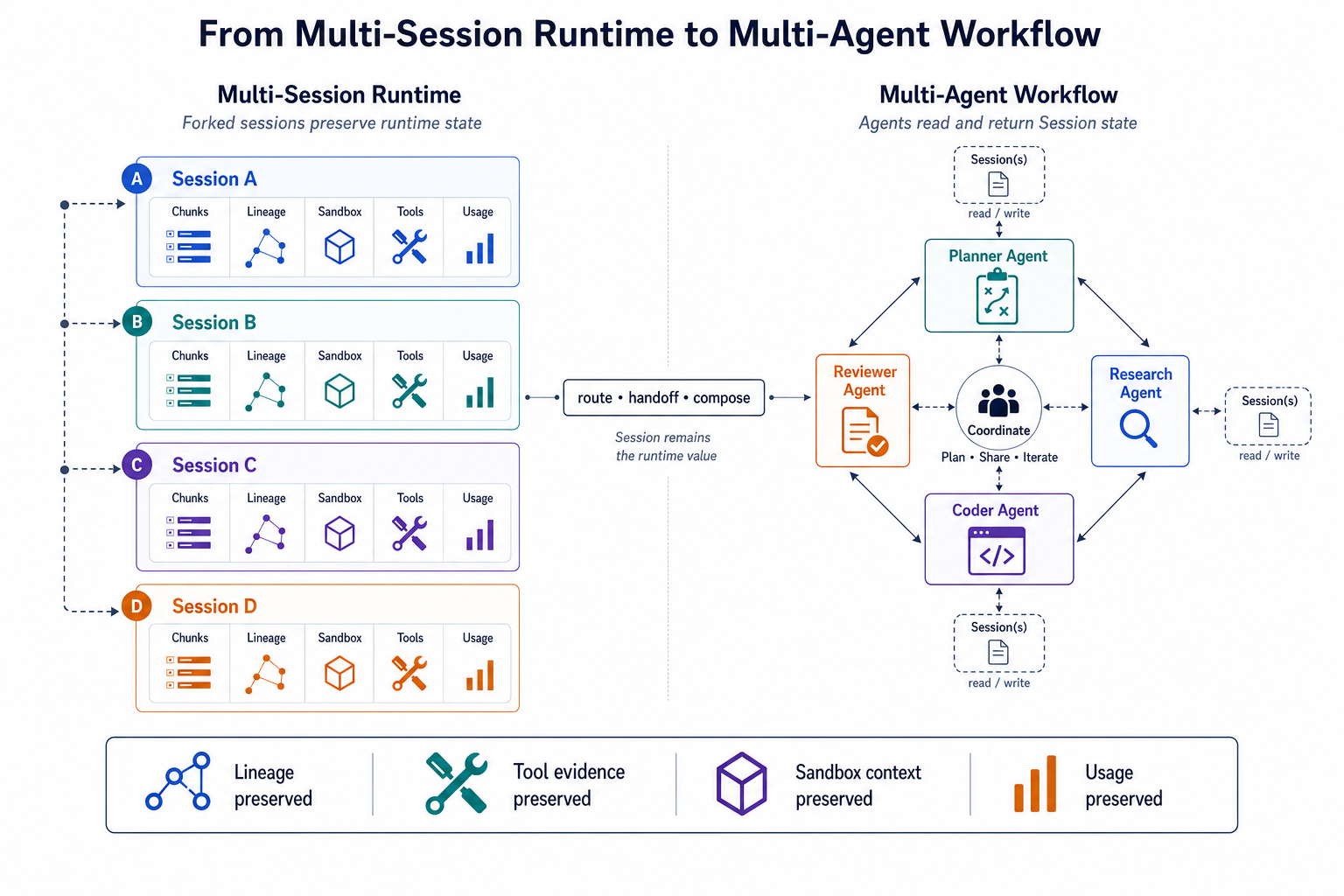}
\caption{Multi-session runtime and multi-agent workflow share the same boundary:
agents route, hand off, and compose work by reading and returning
\texttt{Session} state rather than introducing a second runtime object.}
\label{fig:multi-session-to-agent-workflow}
\end{figure}

\begin{table}[!htbp]
\vspace{0.1em}
\centering
\small
\caption{Multi-agent composition without introducing a second runtime state.}
\label{tab:compact-multi-agent-design}
\begin{tabularx}{\linewidth}{@{}>{\raggedright\arraybackslash}p{0.22\linewidth}>{\raggedright\arraybackslash}X>{\raggedright\arraybackslash}X@{}}
\toprule
Pattern & How it composes & What stays inspectable \\
\midrule
One agent, many sessions & The same agent parameters are applied to fresh, forked, resumed, or sandbox-bound sessions. & Prompt layer stays stable while conversation, placement, lineage, and usage live on the session. \\
Many agents, one state & Specialist agents each consume and return \texttt{Session}; workflows pass the returned value forward. & Intermediate state can be inspected after planning, tool use, compression, QA, or final synthesis. \\
Nested workflows & A child workflow hides internal agent structure behind \texttt{forward(session)}. & Parent workflows see one returned session instead of a private orchestration format. \\
Control surfaces & Chunks, tool results, lineage fields, usage counters, persistence files, and callbacks remain tied to the session. & Failures, budget crossings, branch provenance, and interrupted runs become reviewable artifacts. \\
\bottomrule
\end{tabularx}
\vspace{0.1em}
\end{table}

\section{Implementation Milestones}
\label{sec:milestones}

OpenRath is a working implementation, not an architecture sketch. It is
distributed as a Python package whose modules realize the objects of the
preceding sections: a session core; an execution-backend layer; the flow layer
of tools, agents, workflows, and compressors; an LLM-provider layer; and
persistence with lineage export. This report is written against an audited
snapshot of that codebase, which we treat as adequate for technical review
rather than as a tagged archival release.
Table~\ref{tab:implementation-claim-status} records which surfaces the current
implementation substantiates and where its claims are deliberately bounded.

\begin{table}[!htbp]
\vspace{0.15em}
\centering
\small
\caption{Implementation surface and claim status.}
\label{tab:implementation-claim-status}
\begin{tabularx}{\linewidth}{@{}>{\raggedright\arraybackslash}p{0.22\linewidth}>{\raggedright\arraybackslash}X@{}}
\toprule
Surface & Status \\
\midrule
Session core & Implemented runtime value: ordered chunks, branching via \texttt{fork}, \texttt{detach}, and \texttt{merge}, usage accounting, and JSONL lineage export, exercised by focused tests. \\
Backend placement & Local execution implemented and verified; OpenSandbox available as an optional backend, unconfigured in this environment. \\
Tool layer & Implemented boundary: model-visible schemas with backend-dispatched side effects returned to the session, with custom-tool and MCP examples. \\
Agent and workflow & Implemented composition surface over the uniform \texttt{Session -> Session} contract, including a scripted multi-stage workflow. \\
Provider layer & Live-inference prerequisites in place; model quality and provider runs are out of scope for this report. \\
Memory plane & Intended runtime plane, not yet substantiated by a local module with examples and tests (see below). \\
Examples & A worked example set spanning lineage, backends, tools, streaming, usage, and multi-agent workflows. \\
\bottomrule
\end{tabularx}
\vspace{0.15em}
\end{table}

The substantiated claims are deterministic and local. The session core---its
ordered chunks, branching operations, usage accounting, and JSONL lineage
export---is exercised by focused tests, as are local sandbox placement and the
tool-dispatch path; tool and workflow behavior is further demonstrated by
custom-tool, MCP, and scripted-workflow examples. The optional OpenSandbox
backend and the LLM-provider layer are present but environment- or
provider-dependent, and live model quality lies outside the scope of this report.

What the milestone demonstrates is not any individual surface but that they
share a single object model. \texttt{Session} is the value that flows; backends
determine where code and tools execute; tool calls enter the session as
structured events rather than disappearing into an executor log; agents and
workflows transform sessions without maintaining private transcript formats; and
persistence with lineage export renders the resulting state inspectable outside
the running process. This is the minimum structure an agent runtime requires to
support branching, audit, and replay without rebuilding a separate observability
system for each application.

\section{Release Evidence and Evaluation Protocol}
\label{sec:release-evidence}

The release evaluation is audit-first. It scopes evidence to runtime claims
rather than leaderboard claims: whether OpenRath's runtime claims are backed
by rebuildable evidence packets, whether each packet states its own boundary,
and whether every visible claim is mapped into a claim ledger.

\begin{figure}[!htbp]
\centering
\includegraphics[width=0.92\linewidth]{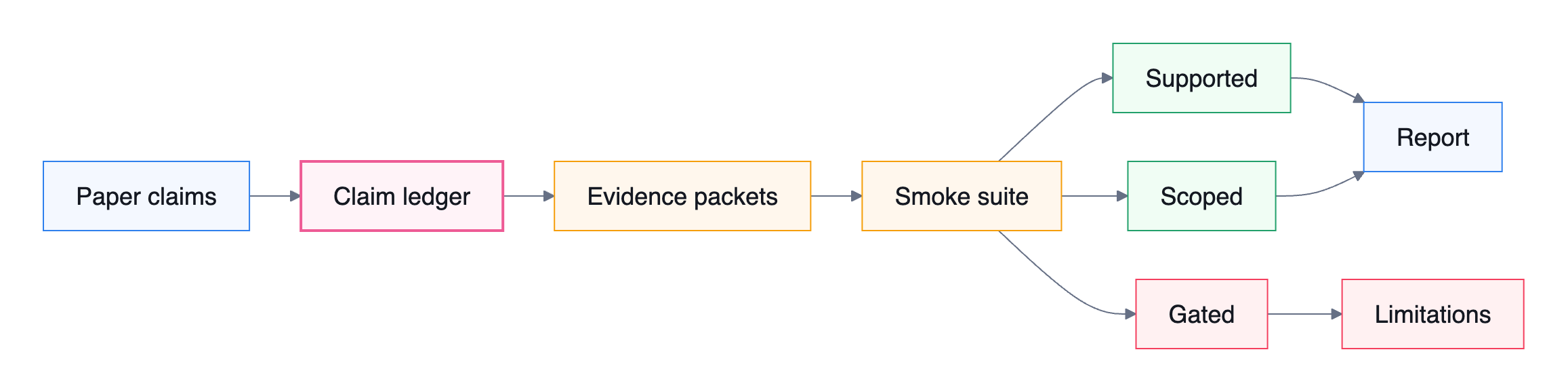}
\caption{Claim-to-evidence protocol: report claims pass through a ledger,
evidence packets, and a smoke suite before becoming supported text, scoped text,
or explicit limitations.}
\label{fig:evidence-protocol}
\end{figure}

The claim ledger is the reviewer-facing contract. It currently classifies ten
claims: five supported by operational packets, one partially supported, one
supported only for prerequisites, one bibliography-backed positioning claim,
one layout smoke claim, and one evidence-gated claim. This keeps the report honest
without making the paper read like an internal backlog: supported claims stay
in the main thesis; evidence-gated claims stay visible and bounded, and framing claims
are separated from empirical evidence.

An evidence packet is deliberately small. It contains the command that produced
the run, a manifest, source and environment metadata, any session JSONL or
tool logs, the generated output artifact, and a short summary of what the
packet does and does not prove. This shape is easier to review than an informal
run transcript because it gives a reader a direct path from paper claim to
reproducible artifact. It also gives maintainers a practical release gate: a
claim can move from evidence-gated to supported only when the corresponding
packet runs under the documented environment.

\begin{table}[!htbp]
\vspace{0.1em}
\centering
\small
\caption{Current release evidence protocol.}
\label{tab:compact-release-evidence-protocol}
\begin{tabularx}{\linewidth}{@{}>{\raggedright\arraybackslash}p{0.22\linewidth}>{\raggedright\arraybackslash}X>{\raggedright\arraybackslash}X@{}}
\toprule
Runtime claim & Current packet & Scope boundary \\
\midrule
Session lineage is inspectable. & \texttt{lineage\_export}: pass, deterministic. & Proves exported branch metadata, not branching quality. \\
Tool placement is auditable. & \texttt{local\_sandbox}: pass; \texttt{opensandbox\_optional}: skip. & Proves local placement evidence, not OpenSandbox parity. \\
Workflows compose session state. & \texttt{workflow\_transcript}: pass, deterministic. & Proves composition shape, not live agent quality. \\
Implementation contracts hold for the focused subset. & \texttt{pytest\_report}: pass. & Does not cover every live integration. \\
Provider prerequisites can be disclosed safely. & \texttt{live\_provider\_manifest}: pass, redacted. & Does not execute live inference. \\
Memory is scoped as a session-visible plane. & \texttt{memory\_local}: skip. & Evidence-gated until source anchors exist. \\
Claim scope is tracked explicitly. & \texttt{claim\_ledger}: pass, ten claims. & One evidence-gated claim remains: \texttt{memory\_runtime\_plane}. \\
Report layout is reviewable. & \texttt{visual\_qa} and \texttt{layout\_audit}: pass. & Visual smoke, not final human design approval. \\
\bottomrule
\end{tabularx}
\vspace{0.1em}
\end{table}

This packet-first evaluation style is especially appropriate before broad
benchmarks. Benchmarks such as curated coding suites~\cite{swebench-verified}
and broad general-assistant evaluations~\cite{gaia} are valuable, but they
combine runtime semantics with model choice, prompt design, environment setup,
reviewer scoring, and task distribution. For a runtime
report, the first question is narrower: can the system preserve and expose the
state needed to make those later evaluations meaningful? OpenRath's current
evidence protocol answers that narrower question before attempting comparative
leaderboard claims.

The baseline and metric design follows the same principle. Follow-on comparisons
should be organized by runtime shape rather than brand name: single-agent loop,
multi-agent shared transcript, workflow/DAG runner, notebook or script
baseline, sandboxed tool agent, and memory/RAG baseline. Metrics should track
runtime correctness, provenance coverage, replayability, backend portability,
efficiency, task quality, and control/safety events. Those metrics become
results only when a runner emits comparable packets for OpenRath and each
baseline.

\section{Limitations}

The limitations are stated as scope boundaries rather than apologies: they
delimit what the report does and does not claim. The report substantiates a
\texttt{Session}-centered runtime object with deterministic evidence for a
narrow set of runtime claims. It does not claim broad benchmark superiority, a
verified local-memory implementation, OpenSandbox availability, fully
reproducible live-model outputs, or any safety property.
Table~\ref{tab:scoped-limitations} records each boundary, the current posture,
and what a stronger claim would require.

\begin{table}[H]
\vspace{0.1em}
\centering
\small
\caption{Scoped limitations for the current technical report.}
\label{tab:scoped-limitations}
\begin{tabularx}{\linewidth}{@{}>{\raggedright\arraybackslash}p{0.18\linewidth}>{\raggedright\arraybackslash}X>{\raggedright\arraybackslash}X@{}}
\toprule
Boundary & Current posture & Required before a stronger claim \\
\midrule
Benchmarking & A deterministic smoke runner and evidence packets, not a broad baseline/metric benchmark. & Pinned workloads, baseline adapters, metrics, live-provider runs, and reviewer-scored artifacts. \\
Backend parity & Local-backend evidence passes; OpenSandbox is an explicit optional skip in this environment. & A backend capability matrix, a configured OpenSandbox packet, and documented mount, export, and failure behavior. \\
Memory & Memory remains an intended runtime plane; \texttt{memory\_local} is evidence-gated because source anchors are absent. & Restored local-memory APIs, examples, and tests, then recall/commit quality evaluated separately from implementation existence. \\
Multi-agent control & Sessions expose branch, merge, tool, and lineage evidence, but do not constitute a policy layer. & Role permissions, tool authority, memory-commit gates, merge policy, and human-review requirements. \\
Safety & No safety property is claimed; tool use and interactive environments enlarge the attack surface, including the data/instruction confusion exploited by indirect prompt injection~\cite{indirect-prompt-injection}. & Evaluation against agent-, web-, and embodied-safety benchmarks~\cite{agent-safetybench,safearena,safeagentbench}, plus tool-authority limits and human-review gates. \\
Reproducibility & Packets support inspection and no-key replay for deterministic claims; live outputs remain provider- and environment-dependent. & Pinned source snapshots, provider manifests, sandbox images, cached external payloads, and disclosure of missing artifacts. \\
\bottomrule
\end{tabularx}
\vspace{0.1em}
\end{table}

These boundaries are part of the release argument: they separate implemented
runtime semantics from optional integrations, follow-on evaluation, and risks
the report does not address. An item should leave this table only when a
supporting evidence packet exists and the claim ledger maps the corresponding
text to that artifact.

\section{Conclusion}

OpenRath's contribution is deliberately narrow: it makes the state that agents
operate on explicit. A multi-agent system is not only a prompt graph, a tool
registry, a trace stream, or a benchmark harness. It is a runtime in which
conversation chunks, branch lineage, sandbox placement, tool effects, memory
interactions, usage, artifacts, and replay evidence must remain connected.

\texttt{Session} is OpenRath's proposed boundary for that runtime state. The
programming model is compact because every major component either transforms a
\texttt{Session}, annotates it, dispatches work through its placement, or emits
evidence that can be inspected after the run. Because that evidence lives in the
value the program already passes around rather than in a side channel
reconstructed afterward, it stays available exactly when a reviewer needs it.
This makes OpenRath complementary to graph runtimes, tracing SDKs, tool
protocols, sandbox providers, and real-environment benchmarks rather than a
replacement for them.

The current technical report therefore makes a scoped claim: deterministic
runtime behavior can be reviewed through release packets today, while broader
quality comparisons, memory-quality evaluation, and live-provider results
belong in follow-on benchmark artifacts. The durable thesis is that reliable
agent systems need a first-class runtime value, and OpenRath makes
\texttt{Session} that value.

That thesis is also the practical standard for the next iteration of the
project. New capabilities should enter the report only when they preserve the
same boundary: they should transform a \texttt{Session}, attach evidence to a
\texttt{Session}, or expose a backend effect through a \texttt{Session}. This
keeps the system from becoming a collection of hidden side channels. It also
keeps the report honest: implementation milestones, case studies, and
evaluation claims can expand without changing the core argument that agent
systems become easier to compose, debug, review, and evaluate when their
runtime state is explicit.

If the last decade of deep learning made the tensor the value a network is
built around, the next generation of agent systems needs the same move: a
single runtime value that everything reads, transforms, and explains. OpenRath
proposes that value is the \texttt{Session}.

\clearpage
\section*{References}
\renewcommand{\bibsection}{}
\bibliographystyle{unsrtnat}
\bibliography{references}

\clearpage
\appendix
\section*{Appendix}

\section{Case Studies}
\label{app:case-studies}

The case studies are used as scoped applicability arguments rather than benchmark results. Their purpose is to identify the kinds of workloads for which a \texttt{Session}-centered runtime model is intended, and to relate those workloads to the deterministic evidence already provided by the release packet suite. The current packets cover the runtime core: lineage export, local sandbox execution, scripted workflow composition, focused implementation tests, and explicit skips for evidence-gated memory and optional OpenSandbox support. The role of the case studies is therefore to connect these audited runtime claims to realistic workload patterns, without converting qualitative applicability into quantitative performance claims.

\begin{table}[!htbp]
\vspace{0.1em}
\centering
\small
\caption{Case-study coverage kept in the main report. Each row states what the workload demonstrates, not a measured leaderboard result.}
\label{tab:case-study-coverage}
\begin{tabularx}{\linewidth}{@{}>{\raggedright\arraybackslash}p{0.22\linewidth}>{\raggedright\arraybackslash}X>{\raggedright\arraybackslash}X@{}}
\toprule
Workload & Runtime point & Current evidence status \\
\midrule
Repository editing & Multi-role planning, sandbox-bound tools, file artifacts, QA review, and lineage around a shared workspace. & Expressiveness coverage; promoted to a result only with a fixed repo seed and measured tests. \\
Research synthesis & Branch-specific context, compression, verifier roles, optional visual tooling, and final style head. & Expressiveness coverage; promoted to a result only with citation-faithfulness and compression-ablation packets. \\
Long-running coding & Persistence, resumability, budget control, context compression, sandbox continuity, and recovery after interruption. & Runtime coverage; promoted to a result only with seeded task packets and recovery measurements. \\
Memory-assisted workflow & Recall/commit should be visible as session operations rather than hidden prompt side effects. & Evidence-gated: audited source lacks current local-memory anchors. \\
Sandbox-isolated execution & Tool effects should carry backend placement, command/file/code logs, and cleanup status. & Local packet passes; optional OpenSandbox packet records a skip. \\
\bottomrule
\end{tabularx}
\vspace{0.1em}
\end{table}

Repository editing and long-running coding provide the most direct motivation for durable and inspectable agent state. These settings include software-engineering benchmarks built from real GitHub issues~\cite{swebench}, agent-computer interfaces for navigating, editing, and testing repositories~\cite{swe-agent}, and company-style digital-worker tasks~\cite{theagentcompany}. Research synthesis motivates branch-specific context, verifier roles, and controlled compression, since different branches may collect, filter, and restyle evidence before a final synthesis is produced. Memory-assisted workflows define the intended persistent-state plane, but remain scoped because the audited source tree does not yet provide current local-memory anchors. Sandbox-isolated execution is the strongest implementation-backed case: the local packet already records command execution, file effects, code execution, and cleanup status.


\end{document}